\documentclass[a4paper]{article}

\usepackage[english]{babel}
\usepackage[utf8x]{inputenc}
\usepackage{amsmath}
\usepackage{graphicx}
\usepackage{hyperref}
\pdfoutput=1

\title{{\bf Bubble visualization in \\ a simulated hydraulic jump}}
\author{\Large Adam M. Witt \\ 
\Large Dr. John S. Gulliver \\ 
\Large Dr. Lian Shen \\
\\
St Anthony Falls Laboratory \\ Department of Civil Engineering \\ 
University of Minnesota, Minneapolis, MN 55414, USA}

\begin{document}
\maketitle

\begin{abstract}
This is a fluid dynamics video of two- and three-dimensional computational fluid dynamics simulations carried out at St. Anthony Falls Laboratory.  A transient hydraulic jump is simulated using OpenFOAM, an open source numerical solver.  A Volume of Fluid numerical method is employed with a realizable k-epsilon turbulence model.  The goal of this research is to model the void fraction and bubble size in a transient hydraulic jump.  This \href{http://ecommons.library.cornell.edu/bitstream/1813/8237/2/LIFTED_H2_EMS T_FUEL.mpg}{fluid dynamics video } depicts the air entrainment characteristics and bubble behavior within a hydraulic jump of Froude number 4.82.      
\end{abstract}

\section{Numerical method}

Two and three-dimensional simulations were conducted assuming an unsteady, turbulent, incompressible flow.  The evolution of the free surface was modeled using interFoam, a VOF solver designed for two immiscible, isothermal, interpenetrating fluids. To discretize the transient terms, a second order Crank-Nicolson scheme was used with an off-centering coefficient of 0.9.  Spatial terms were discretized using a second order accurate Gauss linear differencing scheme.  The time step was limited by a maximum Courant number of 0.5.

The simulations were initialized on a structured, uniform mesh with cells of length 10 mm.  At subsequent refinement levels the cell length was reduced by 50\% in the region of interest.  In this video, two-dimensional simulations are shown with a maximum grid size of 0.625 mm.  This resulted in a computational domain of 1.1E6 cells, which required approximately 8.4E3 computational hours to capture 15s of simulated time at a sampling rate of 20Hz.  Three-dimensional simulations are shown with a maximum grid size of 1.25mm.  This resulted in a computation domain of 1E7 cells, which required approximately 9.8E5 computational hours to capture 15s of simulated time at a sampling rate of 20Hz.

The summation of time-averaged void fraction over depth at four locations was chosen as the principal indicator of grid sensitivity. A summation is representative of the total air entrained into the flow over time, and therefore serves as a gauge of accuracy for both the turbulent processes in the roller and the breakup of large pockets of air.  As these processes become better resolved, modeled void fractions approach the measured values of Murzyn et al. (2005). (2005). At a computation cell size of 1.25mm and a Froude number of 4.82, the two-dimensional model showed a relative error of 0.037 compared to experimental void fraction while the three-dimensional model showed a relative error of 0.013, an improvement of 2.4\% at a factor of 341 times the computational time.

\section{Two-dimensional simulation}

The two-dimensional section of the video highlights several important physical processes that contribute to air entrainment and bubble transport: 
\begin{enumerate}
\item As a supercritical flow encounters a subcritical flow, pockets of air are captured in a recirculating motion.
\item These pockets experience simultaneous breakup and coalescence in the turbulent shear layer.   
\item The largest bubbles can experience multiple breakups, ejecting smaller bubbles that are quickly transported downstream.  
\item If the buoyant force of the bubble is large enough to overcome the advection of the mean flow, it will rise to the free surface and burst into the atmosphere.  
\item Counter rotating vortices form in larger bubbles.
\end{enumerate}

\section{Three-dimensional simulation}

The three-dimensional section of the video displays additional visualizations that highlight air entrainment characteristics and bubble behavior within the hydraulic jump: 
\begin{enumerate}
\item Turbulent ejections at the free surface contribute to air entrainment in the roller.  
\item Large bubbles rise quickly out of the flow in the roller.   
\item A majority of bubbles retain a cylindrical shape in the turbulent shear region.  
\item Bubbles continuously change surface area and volume in the turbulent shear region.
\item The smallest bubbles remain in the flow downstream from the turbulent shear region.        
\end{enumerate}
Isosurface contours about a volume fraction of 0.95, indicating 5\% air, were used to define and visualize the surface of an individual bubble.

\section{References}

Murzyn, F., Mouaze, D., Chaplin, J., 2005. Optical fibre probe measurements of bubbly flow in hydraulic jumps. Int. J. Multiphase Flow 31(1),
141-154.

\end{document}